\documentclass[12pt]{iopart}
\usepackage{graphicx}
\usepackage{color,epsfig}
\usepackage[utf8]{inputenc}
\usepackage{textcomp}
\usepackage{newunicodechar}
\usepackage[left]{lineno}
\usepackage{booktabs}
\usepackage{float}
\usepackage{graphicx}
\usepackage{array}
\newunicodechar{−}{-}
\newunicodechar{θ}{\theta}
\begin{document}
\title{Magnetic Moiré Systems: a review}

\author{Paula Mellado}

\address{Facultad de Ingeniería y Ciencias. Universidad Adolfo Ibáñez, Santiago, Chile}

\ead{paula.mellado@uai.cl}

\begin{abstract}
This review synthesizes recent advancements in the study of moiré magnetism. This emerging field, at the intersection of twistronics, topology, and strongly correlated systems, explores novel phenomena that arise when moiré potentials influence the magnetism of two-dimensional systems. The manuscript presents recent advances highlighting the interfacial incongruity as a novel mechanism for regulating the magnetism of two-dimensional materials and for the manifestation of various phenomena in twisted and mismatched magnetic two-dimensional interfaces. The manuscript addresses seminal and recent experimental and theoretical advances associated with both small- and large-period magnetic moiré lattices, including novel magnetic phases, low-energy and topological magnetic excitations, magnetic and electronic transport, optical properties, phase transitions, and prospective applications of these materials. Moiré magnetism signifies a promising frontier for manipulating complex quantum states in quantum matter. The ongoing advances in this field are poised to impact condensed matter physics, materials science, and quantum information science.
\end{abstract}
\section{Introduction}
Two-dimensional (2D) magnetic systems differ from three-dimensional (3D) and one-dimensional (1D) counterparts due to 1) enhanced quantum fluctuations from reduced dimensionality, 2) absence of long-range magnetic order without magnetic anisotropy \cite{wei2020emerging,lines1969magnetism,mattis2006theory}, and 3) finite-temperature phase transitions driven by topological magnetic defects. In 2D systems, the interplay between magnetic and electric degrees of freedom manifests itself in intricate and powerful ways, influenced by both intrinsic and external fields \cite{wei2020emerging, lines1969magnetism, mattis2006theory}. This dynamic interaction presents exceptional opportunities for fine-tuning correlations in natural and artificial magnetic materials, posing significant prospects and formidable challenges.
For instance, the presence of magnetic frustration, especially in structures like the Kagome or triangular lattices \cite{ramirez2001geometrical,vojta2018frustration,starykh2015unusual}, can give rise to spin-liquid phases  \cite{broholm2020quantum,savary2016quantum}. These phases exhibit fractional excitations, long-range quantum entanglement in their ground states, topologically protected transport channels, and the potential for high-temperature superconductivity upon doping \cite{lee2007high,jiang2021high}.

By overlaying two atomic layers with a slight lattice misalignment or a small rotation angle, a moiré superlattice is formed \cite{hermann2012periodic}, resulting in altered properties compared to the parent materials. These moiré materials have advanced the study and engineering of strongly correlated phenomena and topological systems in lower dimensions \cite{andrei2021marvels}. 

Research activity associated with magnetism in moiré superlattices is relatively new. The most extensively examined example of magnetism emerging from twisting can be traced back to the study of twisted bilayer graphene (TBG) \cite{gonzalez2017electrically,andrei2020graphene,sharpe2019emergent,liu2016quantum,liu2016quantum,wang2023diverse,serlin2020intrinsic,tao2024valley}. In a 2017 theoretical investigation, it was proposed \cite{gonzalez2017electrically} that interaction effects triggered by twisting could lead to antiferromagnetic (AF) or ferromagnetic (FM) polarization of very localized states near AA-stacked areas in twisted graphene bilayers, depending on the small twist angles and electrical bias between layers. The FM-polarized AA regions under bias developed a spiral magnetic ordering as a result of frustrated AF exchange in the absence of spin-orbit coupling. This work set the stage for exploring vdW twisted bilayers as an alternative to control magnetism and study magnetic frustration using electric fields and twists.

In systems featuring magnetic moiré patterns, the \emph{moiré potential} denotes the periodic potential landscape arising from the intrinsic spatial modulation of the moiré superlattice \cite{shabani2021deep}. The moiré-induced periodic potential landscape induces a periodic modulation of magnetic exchange interactions. This results in spatially varying magnetic textures and effective spin-orbit interactions, which create spatial anisotropies that lead to unique energy levels and localized quantum states, manifested as distinct spectral features that enable the realization of a range of novel quantum phenomena \cite{he2021moire,yang2023moire} and the manifestation of exotic phases of matter \cite{mentink2017manipulating,zaliznyak2003heisenberg}. These phases have a considerable impact on the magnetic and electric transport, as well as the optical properties of the system \cite{yang2023moire}. 

The confinement imposed by the moiré pattern can be controlled through various external parameters, including the twist angle between layers, doping, layer stacking, strain, and external electromagnetic fields \cite{he2021moire}. In moiré materials with magnetic order, external magnetic fields enable the adjustment of excitonic and magnetic properties, promoting correlated and topological states of matter with implications for spintronics and quantum optics \cite{shabani2021deep}.

Moiré superlattices consist of homobilayers of the same material or heterobilayers of different materials, often without needing a twist angle for the latter. Magnetic phases arising from moiré superlattices can be induced at the moiré length scale in systems made out of materials that lack intrinsic magnetism (such as in graphene and transition metal dichalcogenides) \cite{ortiz2024transition}, and in systems with intrinsic atomic-scale magnetism (such as the 2D van der Waals magnets, members of the family of chromium trihalides) \cite{soriano2020magnetic}.

Recent developments in layered magnets have broadened the spectrum of two-dimensional (2D) materials, facilitating the study of intrinsic spin properties at the atomic scale \cite{burch2018magnetism}. Over the past five years, the emergence of moiré superlattices has opened up new avenues for modulating spin-dependent phenomena and creating magnetic ground states. Theoretical studies support these breakthroughs with forecasts of noncollinear magnetic textures \cite{hejazi2020noncollinear,hu2021competing,hejazi2021heterobilayer}, skyrmion phases, as a manifestation of topological magnetism \cite{tong2018skyrmions,hu2023magnetic}, the design of moiré magnon bands \cite{kim2022theory}, and novel one-dimensional magnon modes \cite{wang2020stacking}.

Moiré magnets can adjust interlayer magnetic interactions over moiré-length scales, thus aiding in the stabilization of unique spin configurations. This behavior is exemplified by twisted bilayers of $\rm{CrI_3}$ and $\rm{CrBr_3}$, where the moiré pattern allows for tunable interlayer magnetic couplings, ranging from ferromagnetic to antiferromagnetic arrangements. This capability enables the design of magnetic domains and domain walls at the nanoscale \cite{jang2024direct, li2024observation, wu2023coexisting}. The wide variety of van der Waals (vdW) materials offers structural and electronic flexibility, making moiré systems a promising avenue to explore strongly correlated phases and quantum phenomena \cite{debnath2025exploring, mak2022semiconductor}. Moiré-modulated magnetism has been observed in stacked van der Waals (vdW) semiconductors and two-dimensional magnetic materials, such as $\rm CrI_3$ and $\rm CrGeTe_3$. Different stacking configurations significantly influence magnetic and magnetotransport properties, as demonstrated in $\rm Fe_3GeTe_2/In_2Se_3$ \cite{sun2023quantized} and $\rm CrGeTe_3/In_2Se_3$ \cite{xue2020control} heterostructures and in twisted $\rm Fe_3GeTe_2$ homostructures \cite{kim2020observation}. Furthermore, heterostructures combining a magnetic monolayer with a semiconducting layer can exhibit strong proximity effects, enhancing spin transport through the nonmagnetic layer \cite{behura2021moire,tong2019magnetic}.

In this article, we review the current experimental and theoretical developments in the physics of magnetic moiré systems, focusing on the effects of the moiré potential on the emergence of new phases and excitations, as well as its implications for associated transport phenomena. 
The paper is organized as follows. In Section \ref{s2}, we consider moiré-induced magnetic orders and magnetic domains and the phase transitions that arise in these magnetic superlattices. In Section \ref{s3}, we review magnetic excitations such as moiré magnons, stacking magnons, corner edge states, and the non-trivial band topology associated with such bosonic excitations. In Section \ref{s4}, we further introduce the expanded properties of moiré systems to topological spin textures, which involve moiré skyrmions, stacking domain walls, and moiré merons, where moiré-modulated exchange frustration is mainly responsible. Subsequently, in Section \ref{s5}, we review magnon excitons coupling and multiferroic orders in magnetic superlattices. In Section \ref{s6}, we discuss exotic phenomena associated with polarized transport and present promising results regarding quantum phases that arise in magnetic moiré systems, including spin liquids, the quantum Hall effect, and Chern magnets. 
Section \ref{s7} presents some of the implications of these findings and a perspective on the field. Section \ref{s8} is devoted to concluding remarks.
\begin{figure}
\includegraphics[width=\linewidth]{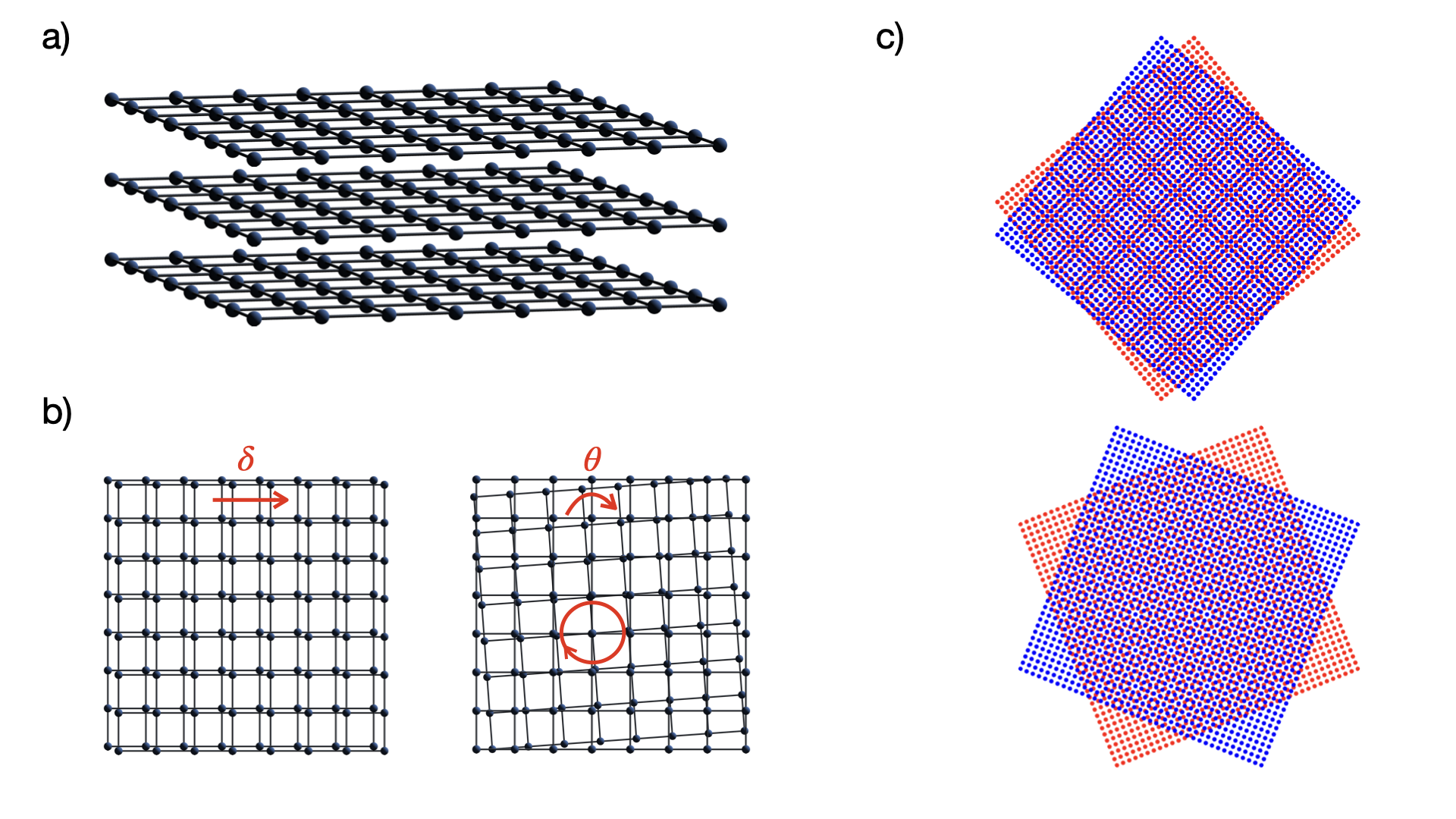}
\caption{Schematic representation of a) a trilayer structure composed of three square lattices stacked vertically. Black spheres represent atoms. b) Two illustrative methods for creating a moiré pattern in a bilayer system: by introducing a relative in-plane mismatch $\delta$ between the layers (left) and by applying a relative twist between the layers at an angle $\theta$ to an orthogonal axis (right). c) Examples of two moiré patterns that emerge when two square lattices are rotated at angles of $\rm\theta=10^\circ$ (top) and $\rm\theta=39^\circ$ (bottom).}
\label{fig1}
\end{figure}
\section{Moiré-induced Magnetic Orders\label{s2}}
\subsection{Magnetism induced at the moiré length scale versus Moiré-modulated intrinsic atomic-scale magnetism} 

Moiré magnets induced at the moiré length scale are systems made out of individual layers that are non-magnetic. The magnetism in these systems arises due to strong electron-electron correlations in the flat bands formed by the moiré potential.

The effective Hamiltonian for such a system can be written as follows 
\begin{equation}
    \mathcal{H}_{\rm{eff}} = \mathcal{H}_{0} + \mathcal{H}_{\rm{moire}} + \mathcal{H}_{\rm{int}}
\end{equation}
where $H_0=\sum_{k,\sigma}\epsilon(k)c_{k,\sigma}^\dagger c_{k,\sigma}$ is the kinetic energy term of the electrons in the individual non-magnetic layers, $\mathcal{H}_{\rm{moire}}$ arises from the periodic potential due to twist/lattice mismatch, which alters the electronic band structure and can lead to flat bands. A spatially varying potential giving rise to this term is of the form $V(\mathbf{r}) = \sum_{\mathbf{G}_m} V_{\mathbf{G}_m} e^{i \mathbf{G}_m \cdot \mathbf{r}}$, where $\mathbf{G}_m$ are moiré reciprocal lattice vectors.  In non-magnetic systems, the last term can be $\mathcal{H}_{\rm{int}}=U \sum_{i} n_{i \uparrow} n_{i \downarrow}$ and describes the Coulomb repulsion, U, between electrons. When the moiré potential flattens the electronic bands by reducing the kinetic energy, the kinetic energy scale, W, becomes comparable to or smaller than the interaction energy scale U ($W\ll U$), and the system is driven into a strongly correlated regime \cite{fazekas1999lecture,gonzalez2017electrically}. In such a case, electrons minimize their interaction energy by ordering their spins, leading to spontaneous symmetry breaking and the emergence of magnetic order at the moiré scale. Therefore, though the magnetism is not present in the individual layers, it is induced by the moiré potential. The order parameter for this induced magnetism is directly tied to the moiré periodicity.

In Moiré-modulated intrinsic atomic-scale magnetism, at least one layer is an intrinsic 2D magnet, and the moiré superlattice modulates the pre-existing atomic-scale magnetic interactions. The effective Hamiltonian for such systems consists of a spin Hamiltonian, which can include various magnetic interactions, which are now modulated by the moiré potential \cite{hejazi2020noncollinear}
\begin{equation}
  \mathcal{H}_{\rm{Spin}} = \sum_{\langle i,j \rangle} J_{ij}(\mathbf{r}_m) \mathbf{S}_i \cdot \mathbf{S}_j + \sum_i K_i(\mathbf{r}_m)(S_i^z)^2 + \sum_{\langle i,j \rangle} \mathbf{D}_{ij}(\mathbf{r}_m) \cdot (\mathbf{S}_i \times \mathbf{S}_j) + \ldots
\end{equation}
where $S_i$ denotes the atomic spin operator at site i, $J_{ij}(\mathbf{r}_m)$ is the exchange coupling between spins i and j. In moiré systems, this exchange interaction becomes spatially modulated at the moiré length scale due to the change in the local stacking configuration within the moiré unit cell \cite{li2022free}. $K_i(\mathbf{r}_m)$ is the magnetic anisotropy term that also becomes spatially modulated, meaning the preferred direction of magnetization can vary across the moiré superlattice. The term $\mathbf{D}_{ij}(\mathbf{r}_m)$ denotes the Dzyaloshinskii-Moriya interaction vector. This term, which favors chiral spin textures, often arises due to broken inversion symmetry at the interface between layers. The moiré potential can create regions where inversion symmetry is locally broken, thus inducing or modulating $\mathbf{D}_{ij}(\mathbf{r}_m)$.  Therefore, the moiré potential modulates these intrinsic atomic-scale magnetic interactions, creating a periodic landscape where the strengths and types of magnetic couplings vary, leading to competing magnetic phases, the emergence of non-collinear spin textures, and the tunability of magnetic properties.

Thus, the distinction between induced magnetism and moiré-modulated intrinsic magnetism lies in the origin of the magnetic order parameter. In the first case, the moiré potential is the primary cause for the appearance of the magnetic order parameter. In the moiré-modulated intrinsic magnetism, the individual magnetic atoms in the layers possess an intrinsic magnetic moment, and the moiré potential acts to reconfigure this pre-existing atomic-scale magnetism.

\subsection{2D magnetic materials}
Chromium trihalides ($\rm{CrX_3}$, with X = Cl, Br, I) serve as archetypal examples among 2D magnetic materials due to their versatile magnetic characteristics and structural versatility \cite{huang2017layer,chen2019direct}. These compounds realize ferromagnetic intralayer order, with out-of-plane easy-axis anisotropy as reported in $\rm{CrBr_3}$ and $\rm{CrI_3}$, and in-plane anisotropy as indicated in $\rm{CrCl_3}$. Bulk $\rm{CrX_3}$ crystals crystallize in either rhombohedral or monoclinic phases, depending on temperature, each displaying unique stacking sequences along the crystallographic $c$ axis \cite{liu2016exfoliating}.

Other layered van der Waals (vdW) materials include $\rm{CrBrS}$, $\rm{MnSe}$, $\rm{MnBi_2Te_4}$, and $\rm Fe_3GeTe_2$ with the first two exhibiting interlayer antiferromagnetic coupling and the third a ferromagnetic interaction \cite{kim2020observation}. Transition-metal phosphorous chalcogenides such as $\rm{MPS_3}$ and $\rm{MPSe_3}$ (M = Mn, Fe, Cr), as well as $\rm{FeTe}$ and $\rm{CrTe_3}$, further enrich the diversity of 2D magnetism. The magnetic and structural properties of these materials can be tuned via external perturbations, such as pressure, electric fields, layer sliding, and twisting \cite{kim2020observation}.

The naturally occurring few-layer $\rm CrI_3$ material realizes the monoclinic stacking. $\rm CrI_3$ has an easy out-of-plane axis for the ferromagnetic intralayer exchange coupling. As a consequence, its long-range magnetic order survives up to the atomic layer limit \cite{wu2019strain,webster2018strain,xie2023evidence}. Experiments and calculations have demonstrated that monoclinic interlayer stacking yields antiferromagnetic (AF) interlayer exchange coupling, whereas rhombohedral interlayer atomic registry leads to ferromagnetic (FM) interlayer interaction \cite{li2019pressure,soriano2019interplay}. 
\subsection{Homostructures and heterostructures comprised of natural materials}
In 2021, Song et al. \cite{song2021direct} constructed twisted 2D chromium triiodide structures and observed emerging magnetic textures, which were not present in individual layers. Single-spin quantum magnetometry experiments revealed moiré magnetism in nanoscale domains and periodic magnetization patterns, demonstrating the coexistence of antiferromagnetic and ferromagnetic domains in a disorder-like pattern. In samples of twisted double-trilayers $\rm CrI_3$, periodic patterns in AF and FM domains matched calculated magnetic structures from interlayer exchange interactions in $\rm CrI_3$ moiré superlattices \cite{song2021direct}.

On the theory front, first-principles simulations have shown that modifying the stacking arrangement in bilayer $\rm{CrX_3}$, the interlayer exchange interaction can be switched from antiferromagnetic to ferromagnetic \cite{sivadas2018stacking,jiang2019stacking}. Specifically, bilayer $\rm{CrI_3}$ undergoes magnetic phase transitions when the stacking geometry is adjusted between the monoclinic and rhombohedral configurations \cite{li2019pressure,song2019switching,kim2022theory}. These transitions can be triggered by external pressure, which strengthens the interlayer magnetic coupling. Further, in bilayer $\rm{CrI_3}$ systems, the application of hydrostatic pressure induces a transition from antiferromagnetic to ferromagnetic interlayer alignment. In contrast, in trilayer structures, the emergence of mixed magnetic domain configurations has been observed \cite{song2019switching}.

Additional studies found that stacking configurations have an impact on the magnetic states of bilayers \cite{xu2022coexisting, xie2023evidence}. Twisted bilayers of $\rm CrI_3$ feature a moiré superlattice comprised of both monoclinic and rhombohedral regions, leading to mixed antiferromagnetic and ferromagnetic interlayer interactions \cite{jiang2019stacking,soriano2019interplay}. This alternancy in the sign of the interlayer exchange coupling is beneficial when the energy gain from antiferromagnetic domains in monoclinic areas exceeds the expense of creating domain walls, particularly below a threshold twist angle. Moreover, studies using magnetic circular dichroism on bilayers with small twist angles revealed both ferromagnetic and antiferromagnetic phases, which shift to a collinear ferromagnetic state once the twist angle surpasses a critical point \cite{xu2022coexisting,yang2024macroscopic,xie2023evidence}. These ferromagnetic-antiferromagnetic phases in twisted bilayers in $\rm CrI_3$ can be controlled through doping by electrical gating \cite{huang2018electrical}.
 
The study of homostructures, each consisting of distinct two-layer $\rm CrI_3$ units, revealed their superior crystalline and magnetic properties over bilayer $\rm CrI_3$. Indeed, research on twisted double bilayers of $\rm CrI_3$ identified a novel magnetic ground state, unlike those in natural two- or four-layer configurations, as reported in \cite{xie2022twist,xie2023evidence}. These systems included two distinct bicrystals $\rm CrI_3$ stackings, forming a moiré superlattice at the interlayer boundaries. The study \cite{xie2022twist,xie2023evidence} illustrated that the emergent magnetic properties in such homostructures could be characterized as a superposition of two- and four-layer $\rm CrI_3$ states under minimal twist conditions. Yet, at larger twist angles, the magnetic properties resemble those of an isolated two-layer $\rm CrI_3$ system \cite{xie2022twist,xie2023evidence}. At intermediate twist angles, unique net magnetization arises due to spin frustration from competing ferromagnetic and antiferromagnetic interactions within moiré supercells \cite{yang2024macroscopic}. While in twisted double odd-layer $\rm CrI_3$, regions with both zero and nonzero magnetization coexist at small twist angles, twisted double even-layer $\rm CrI_3$ exhibits nonzero magnetization near a critical twist angle of $1.1^{\circ}$ \cite{cheng2023electrically}.

In twisted double bilayers of $\rm CrI_3$, the study by \cite{xie2023evidence} provided definitive evidence of non-collinear spin configurations. The non-collinear spin phases were characterized by abrupt spin-flip transitions \cite{xie2023evidence}. A net magnetization from the collinear spin configurations of twisted double bilayers of $\rm CrI_3$ was also present, indicating the coexistence of domains of non-collinear and collinear spins at twist angles between $0.5^{\circ}$ and $5^{\circ}$. In addition, the onset of magnetization and the softening of non-collinear spins in samples with a $1.1^{\circ}$ twist angle occurred at a critical temperature of 25 K \cite{xie2023evidence}, which is much lower than the Néel temperature of 45 K found in several natural layers of $\rm CrI_3$ \cite{liu2018three}. Coexisting antiferromagnetic and ferromagnetic orders with net magnetization were observed via magneto-optical Kerr-effect microscopy over a range of twist angles. These samples exhibited a nonmonotonic temperature dependence, and voltage-assisted magnetic switching was demonstrated \cite{cheng2023electrically}. These magnetic states were also found in numerical simulations, which show that control through twist angle, temperature, and electrical gating is feasible in these systems \cite{xie2022twist}. Remarkably, the domain structures in twisted double bilayers of $\rm CrI_3$, have been manipulated using electric fields \cite{cheng2023electrically}.

In twisted double trilayers of  $\rm CrI_3$, two distinct magnetic phase transitions with separate critical temperatures were observed \cite{li2024observation} in a moiré supercell,  using single-spin scanning magnetometry. Measurements of temperature-dependent spin fluctuations at coexisting ferromagnetic and antiferromagnetic domains showed that in these systems the Curie temperature exceeds the Néel temperature by $\sim 10$ K. Mean-field calculations linked this phase separation to stacking order modulated interlayer exchange coupling at the twisted interface \cite{li2024observation}.

Van der Waals heterostructures where one of the layers is nonmagnetic provide an effective platform for controlling spin in the nonmagnetic layers via the magnetic proximity effect \cite{hauser1969magnetic,bora2021magnetic,cardoso2023strong}. In 2019, experiments demonstrated this control in monolayer transition metal dichalcogenides on a thin ferromagnet \cite{zhong2017van,seyler2018valley}. First-principles calculations \cite{tong2019magnetic} explored this effect in a heterostructure comprising a magnetic monolayer of $\rm CrI_3$ and a semiconductor monolayer (BA), where the lattice mismatch and angular misalignment create a moiré pattern. The magnetic proximity effect, linked to spin-dependent interlayer interactions \cite{ren2024spin,parkin1993origin}, is influenced by the interlayer atomic registry, modulating the magnetic proximity field laterally \cite{tong2019magnetic}. This moiré-modulated effect results in miniband spin splitting, which is highly dependent on the moiré periodicity and can be adjusted through layer twisting, strain, or controlled by a perpendicular electric field \cite{tong2019magnetic}.

It is important to note that the procedures employed in sample fabrication influence the emergence of magnetic phases in the aforementioned moiré systems, and may be a possible explanation for the stacking disorder and the unresolved magnetic coupling that varies with thickness in $\rm CrI_3$ \cite{schneeloch2024role}. In \cite{jang2024direct}, researchers describe unconventional $120^{\circ}$ twisted faults in $\rm CrI_3$ crystals. Exfoliated samples exhibit vertically twisted domains when their thickness is under 10 nm. The preparation techniques influence both the size and distribution of these domains, and cooling procedures can shift the distribution among them \cite{jang2024direct}.

One of the highlights of moiré magnetism is the close interaction between magnetic degrees of freedom and electric fields. Electrically tunable magnetism in a moiré geometry was examined by studying an electron many-body Hamiltonian model for a twisted bilayer of molybdenum ditelluride R stacked $\rm MoTe_2$ \cite{anderson2023programming}. At zero electric field, a correlated honeycomb ferromagnetic insulator was found near one hole per moiré unit cell with a widely tunable Curie temperature up to 14 K. Applying an electric field switched the system into a half-filled triangular lattice with antiferromagnetic interactions; further doping this layer tuned the antiferromagnetic exchange interaction back to ferromagnetic, implying switchable magnetic exchange interactions \cite{anderson2023programming}.

Helical and spiral chiral orders within non-collinear spin phases are of particular significance in the field of spintronics. Magnetochiral textures typically arise from antisymmetric spin interactions stemming from spin-orbit effects, such as the Dzyaloshinskii-Moriya interaction (DMI). DMI occurs in magnetic materials or heterostructures with broken structural space inversion symmetry, leading to chiral and topological spin textures. Non-local dipolar interactions, similar to DMI, anisotropically link the lattice and spin symmetry, stabilizing chiral orders in thin films and low-dimensional lattices \cite{shindou2013chiral}. Recent numerical simulations reveal that twisted square bilayers of magnetic dipoles with easy-plane anisotropy, and interacting through long-range dipolar coupling \cite{tapia2024chiral}, form chiral magnetic phases. Without a twist, the bilayer enters a zigzag magnetic state, with magnetic dipoles ordering in a pattern of ferromagnetic chains that are antiferromagnetically coupled. The moiré patterns from rotating square lattices alter the zigzag order, resulting in phases with non-collinear magnetic textures that feature chiral motifs, which break time and inversion symmetry. At specific moiré angles, helical and toroidal magnetic orders \cite{tapia2024chiral,ding2021field,zimmermann2014ferroic} arise. Additionally, altering the vertical distance between layers can further modify these phases. Mean-field calculations indicated that dipolar interlayer interaction creates a twist-dependent chiral magnetic field, orthogonal to the zigzag chains, responsible for the internal torques linked to toroidal orders \cite{tapia2024chiral}.

In addition to homostructures and heterostructures composed of natural materials, artificial magnetic superlattices offer an alternative avenue, providing enhanced control over moiré magnetism. This is the case of nanostructured magnetic thin films of yttrium-iron-garnet (YIG) \cite{begum2022magnetic,hu2023magnetic} and cold atomic platforms, such as dipolar bosons, with the latest enabling the precise exploration of frustrated magnetism and interaction-driven states in bosonic systems \cite{kennes2021moire,gonzalez2019cold,he2021moire}.

We conclude this section by reviewing theoretical developments in moiré-modulated intrinsic atomic-scale magnetic structures. In this front, a general methodology for deriving continuum models applicable to incommensurate, twisted, or strained multilayer structures, including interlayer coupling effects, was introduced by Hejazi et al. in 2020 \cite{hejazi2020noncollinear}. This framework was exemplified by its application to a twisted bilayer of two-sublattice Néel antiferromagnets situated on a honeycomb lattice, a situation realized in $\rm MnPS_3$ and $\rm MnPSe_3$, and extends its applicability to honeycomb lattice antiferromagnets exhibiting zigzag magnetic order, as well as honeycomb lattice ferromagnets such as $\rm CrI_3$. Employing this approach, authors predicted that the process of twisting these magnetic materials induces the formation of controllable emergent non-collinear spin textures, even though the parent materials inherently display collinear ordering and encompass a diverse array of magnonic subbands. Some of these predictions were soon verified in $\rm CrI_3$ samples \cite{hejazi2020noncollinear}.
\\

\section{Moiré  low energy excitations \label{s3}}
Bosonic phases in magnetic materials are of significant interest due to their neutral and topologically protected boundary modes, as well as their prospective applications in dissipationless magnonics and spintronics. The essential function of moiré patterns in facilitating emergent many-body excitations in magnets was demonstrated through direct visualization of the dispersion of moiré magnons in a monolayer $\rm CrBr_3$ \cite{ganguli2023visualization}. Moiré magnons possess a dispersion pattern that is correlated with the moiré length scale. Low-temperature scanning tunneling microscopy and spectroscopy \cite{ganguli2023visualization} revealed that moiré magnon excitations arise from the interaction of spin excitations in the monolayer $\rm CrBr_3$ and the moiré pattern arising from the lattice mismatch with the underlying substrate. 
Stacking domain walls, possessing substantially higher energy than magnetic domain walls, induces a modulation in the spin Hamiltonian and establishes a stable framework for one-dimensional magnons \cite{song2021direct}. In transition metal dichalcogenides, electrons can be confined in stacking domain walls, which can be controlled experimentally via strain engineering \cite{edelberg2020tunable}. A stacking domain wall also induces a modulation in the spin Hamiltonian \cite{wang2020stacking}. Further, confined 1D magnons have been proposed to exist in magnetic domain walls \cite{abdul2021domain}.  
These domain walls are naturally realized in moiré superlattices in twisted bilayer magnets with small twist angles \cite{song2021direct}. It has been found that in the case of small-angle twisted bilayer $\rm CrI_3$, all stacking domain walls can host 1D magnons, which have energies lower than those of bulk magnons. Such stacking domain walls and corresponding one-dimensional magnons are interconnected \cite{wang2020stacking}, forming a magnon network that dominates low-energy spin and thermal transport.   The existence of these 1D magnons has been traced back to the Goldstone modes of the spin Hamiltonian \cite{wang2020stacking,song2021direct}.
 
The hallmark of an nth-order magnon topological insulator in d-dimensions is the existence of protected gapless magnon states at its (d-n)-dimensional boundaries \cite{bhowmik2024higher,li2022higher,mook2021chiral,mcclarty2022topological,bhowmik2024higher,zhuo2025topological}. Higher-order magnonic topological insulators generally necessitate a strong Dzyaloshinskii-Moriya interaction, which is a comparatively weak phenomenon in the majority of magnetic materials \cite{kuepferling2023measuring}. Theoretical evidence suggests that moiré magnetic systems may offer a solution to this challenge. Indeed, a Heisenberg spin model for misaligned honeycomb ferromagnetic bilayers with a large twist angle \cite{hua2023magnon} was proposed as an ideal platform for second-order topological magnon insulators in the absence of DMI. In this model, the topology of the bilayer depends on the interlayer exchange coupling \cite{hua2023magnon}. In addition, a second-order topological magnon insulator with topologically protected corner states was theoretically found in the ferromagnetic Heisenberg model on a 2D breathing kagome lattice \cite{sil2020first}. A magnonic quadrupole topological insulator that hosts magnon corner states has been theoretically demonstrated in 2D antiskyrmion crystals \cite{hirosawa2020magnonic}, and a second-order topological insulator with 1D chiral hinge magnons was predicted to be realized in 3D stacked honeycomb magnets \cite{mook2021chiral}. 

Brillouin light scattering spectromicroscopy in artificial moiré systems has detected spin-wave moiré edges and cavity modes in a nanostructured magnetic moiré lattice. This lattice consists of two twisted triangular antidot arrays made on a yttrium iron garnet thin film \cite{wang2023observation}. The detection of these spin-wave moiré edge modes occurs at an optimal twist angle. A selective excitation frequency plays a facilitating role, emphasizing the significant influence of dipolar interactions within the system. At each specific twist angle, the magnetic field functions as an additional degree of freedom for modulating the chiral characteristics of the magnon edge modes. Micromagnetic simulations indicate that these edge modes emerge within the intrinsic magnonic band gap and at the interface between a mini flat band and a propagating magnon branch \cite{wang2023observation}. Theoretical calculations of the Berry curvature associated with magnon-magnon coupling indicate a non-trivial topology for the chiral edge modes, reasserting the significance of dipolar interactions in the manifestation of chiral magnetic degrees of freedom in this system \cite{lucassen2019tuning, tapia2024chiral}.

\section{Topological moire spin textures\label{s4}}
Skyrmions represent some of the most prevalent topological defects in magnetic systems and have been observed or theorized in various configurations, including two-dimensional magnets, magnetic thin films, and material interfaces \cite{finocchio2016magnetic}. Their formation is generally attributed to mechanisms that induce chirality, such as DMI interaction, magnetic frustration, and various forms of magnetic anisotropy. However, their occurrence in insulating materials remains infrequent \cite{finocchio2016magnetic}. 
Moiré superlattices offer fertile ground for tuning magnetic topological textures and magnetoelectric couplings.
For instance, moiré-modulated exchange frustration has been shown to give rise to non-collinear magnetic configurations, including skyrmion lattices and helical or spin spirals \cite{kim2024emergence}. Such chiral phases arise from the periodic modulation of interlayer exchange pathways and DMI interactions in systems with broken inversion symmetry.

Results from a combination of ab initio calculations and atomistic simulations in the twisted bilayer of $\rm CrX_3$ with X spanning all three halide ions,  \cite{akram2021moire}, show that for small twist angles, various skyrmion crystal phases can be stabilized in both $\rm CrI_3$ and $\rm CrBr_3$. However, for large angles, all three systems are ferromagnetic. The non-trivial phases result from the interplay between the interlayer AF coupling in the monoclinic stacking zones of the moiré superlattice, plus the energy expense associated with creating the walls of the AF-FM domain \cite{akram2021moire}.

Multiferroics, materials that exhibit simultaneous electric and magnetic order, have recently provided a new avenue for engineering matter by introducing a new building block which, in the form of twisted multiferroic homobilayers, or heterostructures where multiferroic materials team up with magnetic vdW materials, gives rise to spin textures with novel properties \cite{xue2020control,wu2023coexisting}. Twisting has been shown to induce multiferroicity and skyrmionic patterns in bilayer chromium trihalides ($\rm CrX_3$, X=Br, I, Cl) and transition-metal dichalcogenides (TMD) \cite{fumega2023moire,huang2022ferroelectric}. The moiré pattern drives the emergence of multiferroic order in twisted chromium trihalide bilayers and is absent in aligned multilayers.   Using a combination of spin models and ab initio calculations, \cite{fumega2023moire} demonstrated that a spin texture is generated in the moiré supercell of the twisted system as a consequence of the competition between stacking-dependent interlayer magnetic exchange and magnetic anisotropy. An electric polarization arises in association with such a non-collinear magnetic state due to spin-orbit coupling, leading to the emergence of a local ferroelectric order following the moiré. Among the trihalides, results show that twisted $\rm CrBr_3$ bilayers give rise to the strongest multiferroic order where the emergence of a strong magnetoelectric coupling allows the electric generation and control of magnetic skyrmions \cite{sun2023theoretical}. 

$\rm NiI_2$ is classified as a type-II multiferroic, characterized by ferroelectricity arising from its helical magnetic ordering and strong spin−orbit coupling, which result in magnetoelectric coupling \cite{fumega2023moire}. The multiferroic order in $\rm NiI_2$ can be modulated by external factors, including strain, pressure, substrate engineering, or cobalt substitution \cite{fumega2023moire,song2022evidence}. Twisted bilayer $\rm NiI_2$ facilitates the engineering of topological magnetic textures by enhancing the frustration within the system, thereby yielding a variety of exotic magnetic orders \cite{antao2024electric}. These include $k\pi$ skyrmions, whose exotic properties are driven by the commensurability of the spin spiral wavelength and the moiré length scale. Furthermore, the multiferroic order of twisted $\rm NiI_2$ offers a remarkable level of control over skyrmion lattice states through the application of external electric fields, which allows the transition between different skyrmionic configurations \cite{fumega2023moire,song2022evidence}.

The skyrmion in heterostructures composed of one magnetic monolayer on top of a ferroelectric layer with a slight lattice mismatch arises from the DMI effect introduced by the broken inverse symmetry of the heterostructure. In such multiferroic platforms, ferroelectric polarization can be used to stabilize and manipulate magnetic skyrmions. For example, ferroelectric materials such as $\rm BaTiO_3$ or $\rm In_2Se_3$ can be used to influence the Dzyaloshinskii-Moriya interaction in adjacent magnetic materials, leading to the formation or destruction of skyrmions \cite{jin2024skyrmion,zhang2016magnetic,jiang2017direct}. This control can be achieved by manipulating the ferroelectric polarization, effectively allowing for electric-field-driven switching of topological spin textures. This skyrmion can be easily driven by current, where the ferroelectric layer also offers switchable properties in the moiré potential \cite{fumega2023moire} that may cause the skyrmion to move in an ordered potential well, eliminating the transverse drift caused by the Hall effect \cite{jin2024skyrmion,zhang2016magnetic,jiang2017direct}.

Using first-principles calculations and atomistic spin dynamics simulations, the two-dimensional van der Waals $\rm MnS_2/CuInP_2S_6$ multiferroic moiré hetero-superlattice was engineered, exhibiting tunable skyrmions via magnetoelectric coupling \cite{sun2023quantized}. The lattice mismatch between the monolayers of $\rm MnS_2$ and $\rm CuInP_2S_6$ induces incommensurate moiré patterns, along with modulated magnetic anisotropy and emerging magnetic skyrmions in $\rm MnS_2$ \cite{sun2023quantized}. These magnetic skyrmions are affected by magnetoelectric coupling and can be adjusted by ferroelectric polarization of $\rm CuInP_2S_6$. Moreover, the movement of the skyrmions within the moiré period can be regulated by a pulsed current which varies according to the distinct ferroelectric polarization states of the $\rm CuInP_2S_6$ layer.

\section{Exciton-Magnon Coupling\label{s5}}
Most frequently,  magnetic fields are used to modulate magnons, while strain has been shown to generate and control magnons via magnetostriction \cite{lee1955magnetostriction}. In TMD hetero-bilayers, excitons, arising from bound electron-hole pairs through the Coulomb interaction, are promising for their strong light-matter interactions, spin-valley-coupled optical effects and moiré superlattice physics \cite{scholes2006excitons,sell1967optical,klaproth2023origin}. Integrating magnonic information into systems requires effective magnon-exciton interactions for coupling with optical photons \cite{diederich2023tunable}. For efficient information processing, these interactions must be tunable \cite{freeman1968exciton}. Free excitons coupled to magnons have been reported in both bulk antiferromagnetic $\rm MnPS_3$ and heterostructures composed of semiconducting $\rm MoSe_2$ and antiferromagnetic $\rm MnPSe_3$ \cite{sell1967optical,wang2023exciton}. In heterostructures made out of semiconductors (e.g., CrI$_3$/WSe$_2$), moiré potentials modify the excitonic states and mediate their coupling with collective spin excitations \cite{diederich2023tunable,wang2022light}, presenting a novel approach for optically probing magnetic order and spin dynamics.

The interaction between moiré excitonic states (excitons and trions) and magnetic excitations has been explored experimentally \cite{zhang2022magnon}. Excitons were found to be confined in moiré patterns, where intralayer and interlayer excitons display distinct signatures \cite{zhang2022magnon}. These excitons, modulated by moiré potentials, form ordered arrays that are linked to localized intralayer trion-magnon complexes \cite{zhang2022magnon}.

$\rm CrSBr$ is a layered semiconductor in which individual layers are ferromagnetic. This magnetic material has an easy axis anisotropy in the plane and realizes an AF interlayer coupling \cite{ziebel2024crsbr,liu2024intralayer}. Its exciton energy depends on the magnetic configuration of the interlayer, because the spin alignment controls the hopping between layers and thus the optical bandgap \cite{shi2024giant,sell1967optical}. This strong coupling between exciton resonance and interlayer spin coupling is the underlying mechanism of the recent observation of long-lived magnons \cite{bae2022exciton}. In this work, uniaxial strain was applied to tune the $\rm CrSBr$ magnetic anisotropy and to flip the sign of the interlayer magnetic exchange interaction \cite{cenker2022reversible,liu2024intralayer}. Control over coherent coupling between excitons and both bright and dark magnon modes was demonstrated through the use of external magnetic fields and uniaxial strain, as shown by transient optical reflectivity measurements \cite{diederich2023tunable}.
 
In the case of the moiré heterostructure $\rm WS_2 /WSe_2$, optical excitations can tune the spin-spin interactions between moiré-trapped carriers, resulting in an emergent ferromagnetic phase \cite{jin2019observation,wang2022light}. At one hole per three moiré unit cells, setting the power at the exciton resonance gives rise to a hysteresis loop in the reflective magnetic circular dichroism signal, which is an indication of ferromagnetism in the system \cite{jin2019observation}. The hysteresis loop persists even at charge neutrality. It evolves as the moiré superlattice fills, indicating changes in magnetic ground-state properties and suggesting that photoexcited excitons mediate the exchange coupling between moiré-trapped holes \cite{jin2019observation}.
 
\section{Exotic phenomena in moire superlattices \label{s6}}
Moiré superlattices composed of magnetic materials have been demonstrated to be an outstanding platform for the investigation of unconventional phases of matter \cite{begum2022magnetic,das2025moire,regan2024spin,akif2024theory,debnath2025exploring,wang2023diverse,jin2019observation}.

In the case of the long-sought quantum spin-liquid state, the differences between moiré material physics and the single-band Hubbard model have been recently studied \cite{hu2021competing} to explore the possibility of achieving exotic spin-liquid states aided by moiré potentials. By employing Hartree-Fock theory and strong-coupling expansions, the metallic and insulating states in transition-metal dichalcogenide heterobilayers at one electron per moiré period were examined under various conditions \cite{hu2021competing}. Four magnetic phases and one nonmagnetic state near the metal-insulator transition were found. The results indicate that the ferromagnetic insulating states are influenced by nonlocal moiré-modulated exchange interactions \cite{hu2021competing}. These findings establish a lower bound on the moiré modulation strength required for a metallic-to-insulating transition. They show at the microscopic level that uniaxial strain affects exciton–magnon coupling \cite{sell1967optical} and magnon dispersion \cite{hu2021competing}.

The anomalous Hall effect is a transverse voltage that occurs in ferromagnetic conductors without an external magnetic field, commonly found in materials with strong spin-orbit coupling \cite{liu2016quantum,nagaosa2010anomalous}. A quantized anomalous Hall (QAH) effect requires a system with broken time-reversal symmetry and topologically nontrivial bands \cite{liu2016quantum,nagaosa2010anomalous}. Moiré heterostructures serve efficiently in producing intrinsic Chern magnets functioning as Chern insulators. Within a trilayer heterostructure comprising a $\rm 2H-MoTe_2$ bilayer combined with a $\rm WSe_2$ monolayer \cite{tschirhart2023intrinsic}, a strong electric field generates moiré subbands. When there is one hole per superlattice unit cell, a near QAH effect is observed along with low resistance at zero magnetic field. This is akin to occurrences in AB-stacked $\rm MoTe_2/WSe_2$ bilayers, where magnetic Chern insulators are formed due to Coulomb interaction and Berry curvature \cite{wang2023diverse}.

Inspired by the discoveries of quantum anomalous Hall effects in  $\rm MoTe_2$ heterostructures, a recent work \cite{zhou2025itinerant} has proposed realizing itinerant topological magnons in twisted $\rm MoTe_2$ systems, which could be tunable via a displacement field. The proposed effective model links the nontrivial topology of magnons to electronic bands, where itinerant spin excitons arise from particle-hole bound state splitting caused by the electronic band gap. Experiments could utilize spectral measurements to verify the existence of magnons and spin excitons, while changes in thermal Hall conductance may indicate topological transitions \cite{zhou2025itinerant}.

Stacking magnetic with nonmagnetic semiconducting 2D materials results in a magnetic proximity effect \cite{hauser1969magnetic,tong2019magnetic}, causing spin and valley splitting in nonmagnetic 2D layers \cite{seyler2018valley}. This effect is known in ferromagnetic/graphene systems and is helpful for spin transport devices \cite{zhang2014proximity}. In moiré systems, the magnetic proximity effect becomes tunable, as reported in the $\rm CrI_3$/BAs moiré superlattice \cite{zhang2014proximity}, where BAs display a localized miniband due to the periodic moiré potential. Studies \cite{gong2024spin} on nanoribbon devices in moiré superlattices show that stacking a ferromagnetic monolayer ($\rm CrI_3$) on BAs creates moiré patterns due to lattice mismatch and twist angles. These patterns affect interlayer interactions, magnetic proximity, and interlayer distance. The conductance shows spin-resolved miniband transport at small twist angles, with spin-polarized currents arising from spin-resolved minigaps. In this case, only a finite moiré period is needed for full spin polarization, and moiré-modified interlayer distances allow perpendicular electric fields to direct spin polarization \cite{zhang2014proximity}.

In their 2022 article, \cite{PhysRevX.12.031042}, Mejkal et al. introduced the concept of altermagnetism as a new category of collinear antiferromagnets characterized by the violation of specific spatial symmetries. This results in the spin-splitting of electronic bands, despite the absence of net magnetization, and stems from the breaking of both time-reversal and crystalline symmetries, allowing spin-dependent transport without spin-orbit coupling \cite{PhysRevX.12.031042}. Such combinations yield properties such as the Hall effect, nonrelativistic spin current, and giant thermal transport in crystals \cite{mazin2022altermagnetism}. The emergence of altermagnetism necessitates both the presence of compensated collinear magnetic order and the existence of opposite spin sublattices connected by rotational symmetry \cite{mazin2022altermagnetism}. Theoretical investigations \cite{sheoran2024nonrelativistic,liu2024twisted} demonstrate that twisted bilayer configurations consisting of centrosymmetric antiferromagnetic materials are capable of exhibiting nonrelativistic spin splitting, typically associated with the breaking of inversion or time-reversal symmetry. Notably, even in the absence of significant spin-orbit coupling, the symmetry properties inherent to the twisted bilayer facilitate the emergence of spin splittings. These findings introduce a novel type of spin splitting induced by magnetic texture and symmetry breaking within twisted AF bilayers. It uncovers an inherently nonrelativistic mechanism to eliminate spin degeneracy, thereby broadening the domain of altermagnetism and proposing a platform for spintronic applications that are resilient to the challenges associated with spin-orbit coupling.

The Kondo model, employed in the study of lanthanide and actinide intermetallic compounds, describes antiferromagnetic interactions between localized magnetic moments and conduction electrons, characterized by the formation of heavy fermions, magnetic ordering, and the emergence of unconventional superconductivity \cite{coleman1995simple}. Recently, van der Waals heterostructures have provided an alternative realization of the Kondo model, facilitating the study of emergent magnetism. In these structures, synthetic Kondo lattices are composed of a two-dimensional Mott insulator layer and a metallic or semimetallic layer \cite{das2025moire}. Examples include bilayers with $\rm 1T-TaS_2$ as a Mott insulator and $\rm 2H-TaS_2$ as a metal, and hetero-bilayers $\rm 1T/1H-TaSe_2$ and $\rm 1T/1H-TaS_2$ \cite{guerci2024topological,zhao2023gate,zhao2024emergence}. Effective spin models and Monte Carlo simulations demonstrate that stacking-dependent interlayer Kondo interactions give rise to various magnetic orders, resulting in domains within the moiré unit cell \cite{xie2024kondo,das2025moire}. AB (AA) stacking regions exhibit ferro- (or antiferromagnetic) order over a broad range of parameters. Other regions form ferromagnetic chains coupled antiferromagnetically, with the decay length of the Kondo interaction influencing the extent of the phase \cite{das2025moire}.

As a final note, we emphasize the potential that magnetic moiré systems offer to provide a fresh perspective on fractional excitations, which has attracted significant interest within the broader scientific community, particularly among condensed matter physicists and materials science researchers. In this regard, theoretical propositions indicate that systems incorporating itinerant layers, such as graphene coupled with Kitaev quantum spin liquids (e.g., $\alpha$-$\rm RuCl_3$), might offer a means to explore fractionalized excitations. Within these moiré-modulated heterostructures, the manipulation of the Kitaev exchange is feasible, which could facilitate the stabilization of spin-liquid phases \cite{leeb2021anomalous}. These systems show transport signatures that pinpoint scattering from emergent Majorana fermions and visons, with adjustable scattering rates through gating and Fermi velocity regulation \cite{takagi2019concept,liu2021towards}.
\begin{figure}
\includegraphics[width=1.1\linewidth]{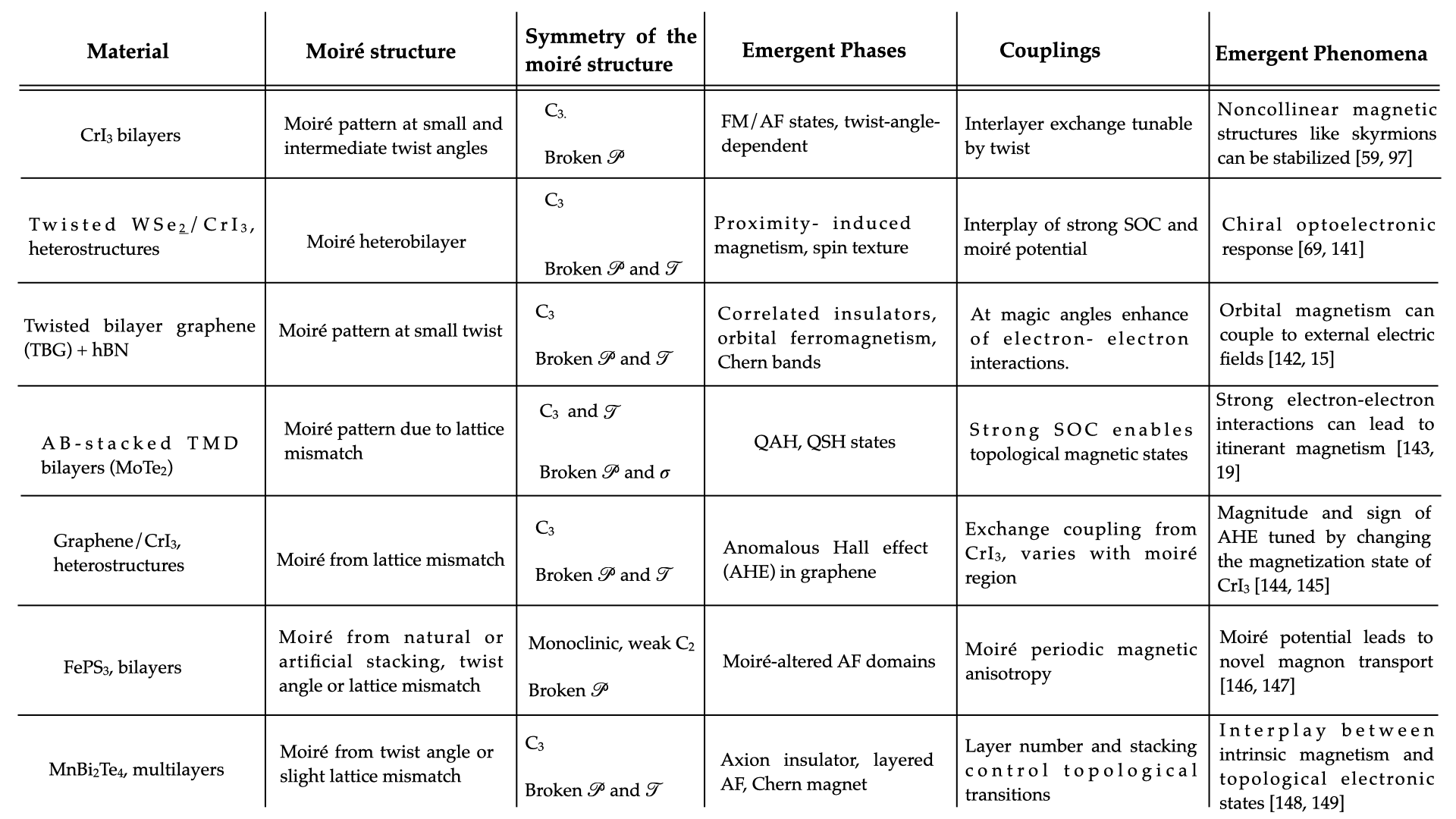}
\caption{Summary table comparing the emergent properties of several magnetic materials that exhibit moiré physics.$\mathcal{P}$, $\sigma$, and $\mathcal{T}$ denote inversion, mirror, and time reversal symmetry, respectively.}
\label{fig:enter-label}
\end{figure}
\section{Perspectives \label{s7}}
In this section, we examine a range of opportunities, challenges, and promising research directions that emerge from this dynamic field of study. From the perspective of practical applications, the finding that moiré fields, akin to antisymmetric interactions, can stabilize skyrmions within designated regions or form skyrmion lattices, even without a marked Dzyaloshinskii-Moriya interaction, is particularly noteworthy. The moiré potential affects the size, shape, and formation of skyrmions, thereby facilitating the engineering and stabilization of skyrmion lattices or skyrmion bubbles under experimental conditions that are considerably more attainable than in traditional approaches \cite{antao2024electric,akram2021moire,tong2018skyrmions,hejazi2021heterobilayer}. In addition, moiré potentials expand the spectrum of opportunities to drive these magnetic topological textures. In twisted bilayers of $\rm CrBr_3$, the manifestation of a moiré pattern can induce a multiferroic state, which may be harnessed to regulate skyrmions via electric fields \cite{huang2022ferroelectric}. Furthermore, the moiré potential facilitates the propulsion of skyrmions via mechanisms such as domain walls anchored by the moiré pattern, which serve as channels for skyrmion movement \cite{shaban2023skyrmion}. These findings are opening up new avenues in the field of spintronics.

Another practical opportunity and significant advantage of moiré systems is that they enable systems to interact in situ with electrostatic gates \cite{anderson2023programming}. This dynamic control allows the manipulation of magnetic phases, phase transitions, and magnetic domains, opening up exciting possibilities for new magnetic memory and logic devices \cite{anderson2023programming}. Moreover, the application of strain to moiré superlattices can further modulate the moiré potential and, consequently, the magnetic couplings, providing an additional mechanism for regulating magnetic properties. The interplay between optical characteristics and magnetism in moiré materials may lead to novel approaches for manipulating light using magnetic fields, or vice versa \cite{wang2022light}.

Promising research directions in moiré magnetic structures emerge from the interaction of magnetism with various quantum phenomena. A fascinating avenue is the connection between moiré magnetism and unconventional superconductivity. Increasing evidence suggests the presence of competing magnetic orders alongside superconducting states in twisted graphene systems. Exploring how moiré patterns might mediate or enhance superconductivity through magnetic fluctuations represents a vital area of research \cite{balents2020superconductivity}. Current studies are investigating how moiré-modulated magnetic order interacts with and can give rise to topological phases of matter, such as the quantum anomalous Hall effect and topological superconductivity, where Majorana zero modes could potentially be realized and manipulated \cite{adak2024tunable,kezilebieke2022moire}.
Additionally, understanding the interplay between electronic excitations, such as excitons, and magnetism in magnetic van der Waals structures is gaining considerable attention \cite{zhou2025itinerant,jin2019observation,diederich2023tunable}. This includes predicting and observing moiré excitons and comprehending how interlayer exchange interactions and external fields modulate these phenomena.

The influence of moiré potentials on spin liquids and their effects on phenomena associated with magnetic frustration represents a largely uncharted territory. Presently, emerging numerical studies are beginning to shed light on the role played by moiré physics in the ground states of twisted bilayers of artificial spin ice lattices \cite{begum2022magnetic}. However, the effect of moiré potentials on the low-energy physics of quantum spin liquids and spin glasses remains a largely unexplored domain. 

Overall, a significant challenge persists in accurately characterizing the complex many-body interactions within the expansive moiré unit cells.
Advances in theoretical modeling, including first-principles density functional theory calculations and continuum field theories, represent new research opportunities and are crucial for predicting new moiré magnetic phases, understanding their underlying mechanisms, and guiding experimental efforts.

\section{Concluding Remarks \label{s8}}
Magnetic moiré systems represent a fertile frontier in the investigation of correlated quantum states and topological excitations within two-dimensional systems, which is crucial for advancing our understanding of quantum materials. The superposition of slightly mismatched atomic lattices yields moiré superlattices that induce spatial modulation of magnetic interactions, resulting in the emergence of novel magnetic phases, including coexisting ferromagnetic and antiferromagnetic orders, noncollinear spin textures, and topological spin configurations. Beyond emergent magnetic orders, low-energy bosonic excitations, such as moiré magnons, edge modes in antidot arrays, and higher-order topological magnon states, have been documented, establishing links with the physics of magnonic topological insulators. These excitations can be controlled through electric fields, pressure, twist angle, or proximity coupling, becoming versatile tools for designing functional quantum devices. Recent research also demonstrates the coupling between moiré-confined excitons and magnons, paving the way for optospintronics and the quantum simulation of many-body phenomena. In multiferroic heterostructures, twisting and lattice mismatch yield novel spin-charge textures and controllable magnetoelectric couplings, enabling the electric-field manipulation of topological defects. On the electronic front, magnetic moiré systems have unveiled exotic phases, including Mott insulators, spin liquids, and anomalous quantum Hall states. Moiré magnetic proximity effects introduce spatially modulated effective magnetic fields within nonmagnetic layers, thereby facilitating the generation of spin-polarized currents and the engineering of Chern insulators. The recent identification of \textit{altermagnetism} in twisted bilayers of centrosymmetric antiferromagnets—even in the absence of spin-orbit coupling—posits a new avenue for the creation of spintronic platforms that are resilient to decoherence. Furthermore, theoretical propositions exploring fractionalized excitations in moiré heterostructures comprising itinerant layers (such as graphene) with Kitaev spin liquids highlight the potential of moiré systems for the investigation of anyonic statistics and topological quantum computation. 

In sum, progress in magnetic moiré systems is transforming the landscape of quantum condensed matter physics, with profound implications for material science, quantum information, and the forthcoming wave of quantum technologies. 
\subsection{Acknowledgments}
The author acknowledges support of Fondo Nacional de Desarrollo Científico y Tecnológico (Fondecyt) under Grant No. 1250122.

\section{References}
\providecommand{\newblock}{}


\begin{thebibliography}{100}
\expandafter\ifx\csname url\endcsname\relax
  \def\url#1{{\tt #1}}\fi
\expandafter\ifx\csname urlprefix\endcsname\relax\def\urlprefix{URL }\fi
\providecommand{\eprint}[2][]{\url{#2}}

\bibitem{wei2020emerging}
Wei S, Liao X, Wang C, Li J, Zhang H, Zeng Y~J, Linghu J, Jin H and Wei Y 2020 {\em 2D Materials\/} {\bf 8} 012005

\bibitem{lines1969magnetism}
Lines M 1969 {\em Journal of Applied Physics\/} {\bf 40} 1352--1358

\bibitem{mattis2006theory}
Mattis D~C 2006 {\em Theory Of magnetism made simple, the: an introduction to physical concepts and to some useful mathematical methods\/} (World Scientific Publishing Company)

\bibitem{ramirez2001geometrical}
Ramirez A 2001 {\em Handbook of magnetic materials\/} {\bf 13} 423--520

\bibitem{vojta2018frustration}
Vojta M 2018 {\em Reports on Progress in Physics\/} {\bf 81} 064501

\bibitem{starykh2015unusual}
Starykh O~A 2015 {\em Reports on Progress in Physics\/} {\bf 78} 052502

\bibitem{broholm2020quantum}
Broholm C, Cava R~J, Kivelson S, Nocera D, Norman M and Senthil T 2020 {\em Science\/} {\bf 367} eaay0668

\bibitem{savary2016quantum}
Savary L and Balents L 2016 {\em Reports on Progress in Physics\/} {\bf 80} 016502

\bibitem{lee2007high}
Lee P~A 2007 {\em Reports on Progress in Physics\/} {\bf 71} 012501

\bibitem{jiang2021high}
Jiang H~C and Kivelson S~A 2021 {\em Physical Review Letters\/} {\bf 127} 097002

\bibitem{hermann2012periodic}
Hermann K 2012 {\em Journal of Physics: Condensed Matter\/} {\bf 24} 314210

\bibitem{andrei2021marvels}
Andrei E~Y, Efetov D~K, Jarillo-Herrero P, MacDonald A~H, Mak K~F, Senthil T, Tutuc E, Yazdani A and Young A~F 2021 {\em Nature Reviews Materials\/} {\bf 6} 201--206

\bibitem{gonzalez2017electrically}
Gonzalez-Arraga L~A, Lado J, Guinea F and San-Jose P 2017 {\em Physical review letters\/} {\bf 119} 107201

\bibitem{andrei2020graphene}
Andrei E~Y and MacDonald A~H 2020 {\em Nature materials\/} {\bf 19} 1265--1275

\bibitem{sharpe2019emergent}
Sharpe A~L, Fox E~J, Barnard A~W, Finney J, Watanabe K, Taniguchi T, Kastner M and Goldhaber-Gordon D 2019 {\em Science\/} {\bf 365} 605--608

\bibitem{liu2016quantum}
Liu C~X, Zhang S~C and Qi X~L 2016 {\em Annual Review of Condensed Matter Physics\/} {\bf 7} 301--321

\bibitem{wang2023diverse}
Wang T, Devakul T, Zaletel M~P and Fu L 2023 {\em arXiv preprint arXiv:2306.02501\/}

\bibitem{serlin2020intrinsic}
Serlin M, Tschirhart C, Polshyn H, Zhang Y, Zhu J, Watanabe K, Taniguchi T, Balents L and Young A 2020 {\em Science\/} {\bf 367} 900--903

\bibitem{tao2024valley}
Tao Z, Shen B, Jiang S, Li T, Li L, Ma L, Zhao W, Hu J, Pistunova K, Watanabe K {\em et~al.\/} 2024 {\em Physical Review X\/} {\bf 14} 011004

\bibitem{shabani2021deep}
Shabani S, Halbertal D, Wu W, Chen M, Liu S, Hone J, Yao W, Basov D~N, Zhu X and Pasupathy A~N 2021 {\em Nature Physics\/} {\bf 17} 720--725

\bibitem{he2021moire}
He F, Zhou Y, Ye Z, Cho S~H, Jeong J, Meng X and Wang Y 2021 {\em ACS nano\/} {\bf 15} 5944--5958

\bibitem{yang2023moire}
Yang B, Li Y, Xiang H, Lin H and Huang B 2023 {\em Nature Computational Science\/} {\bf 3} 314--320

\bibitem{mentink2017manipulating}
Mentink J 2017 {\em Journal of Physics: Condensed Matter\/} {\bf 29} 453001

\bibitem{zaliznyak2003heisenberg}
Zaliznyak I~A 2003 {\em Physical Review B\/} {\bf 68} 134451

\bibitem{ortiz2024transition}
Ortiz~Jimenez V, Pham Y~T~H, Zhou D, Liu M, Nugera F~A, Kalappattil V, Eggers T, Hoang K, Duong D~L, Terrones M {\em et~al.\/} 2024 {\em Advanced Science\/} {\bf 11} 2304792

\bibitem{soriano2020magnetic}
Soriano D, Katsnelson M~I and Fern{\'a}ndez-Rossier J 2020 {\em Nano Letters\/} {\bf 20} 6225--6234

\bibitem{burch2018magnetism}
Burch K~S, Mandrus D and Park J~G 2018 {\em Nature\/} {\bf 563} 47--52

\bibitem{hejazi2020noncollinear}
Hejazi K, Luo Z~X and Balents L 2020 {\em Proceedings of the National Academy of Sciences\/} {\bf 117} 10721--10726

\bibitem{hu2021competing}
Hu N~C and MacDonald A~H 2021 {\em Physical Review B\/} {\bf 104} 214403

\bibitem{hejazi2021heterobilayer}
Hejazi K, Luo Z~X and Balents L 2021 {\em Physical Review B\/} {\bf 104} L100406

\bibitem{tong2018skyrmions}
Tong Q, Liu F, Xiao J and Yao W 2018 {\em Nano letters\/} {\bf 18} 7194--7199

\bibitem{hu2023magnetic}
Hu Y and Du A 2023 {\em Journal of Magnetism and Magnetic Materials\/} {\bf 588} 171374

\bibitem{kim2022theory}
Kim K~M, Kiem D~H, Bednik G, Han M~J and Park M~J 2022 {\em arXiv preprint arXiv:2206.05264\/}

\bibitem{wang2020stacking}
Wang C, Gao Y, Lv H, Xu X and Xiao D 2020 {\em Physical Review Letters\/} {\bf 125} 247201

\bibitem{jang2024direct}
Jang M, Lee S, Cantos-Prieto F, Ko{\v{s}}i{\'c} I, Li Y, McCray A~R, Jung M~H, Yoon J~Y, Boddapati L, Deepak F~L {\em et~al.\/} 2024 {\em Nature communications\/} {\bf 15} 5925

\bibitem{li2024observation}
Li S, Sun Z, McLaughlin N~J, Sharmin A, Agarwal N, Huang M, Sung S~H, Lu H, Yan S, Lei H {\em et~al.\/} 2024 {\em Nature Communications\/} {\bf 15} 5712

\bibitem{wu2023coexisting}
Wu Y, Tong J, Deng L, Luo F, Tian F, Qin G and Zhang X 2023 {\em Nano Letters\/} {\bf 23} 6226--6232

\bibitem{debnath2025exploring}
Debnath S, Dey S and Giri P 2025 {\em ACS Applied Materials \& Interfaces\/}

\bibitem{mak2022semiconductor}
Mak K~F and Shan J 2022 {\em Nature Nanotechnology\/} {\bf 17} 686--695

\bibitem{sun2023quantized}
Sun W, Wang W, Yang C, Li X, Li H, Huang S and Cheng Z 2023 {\em Physical Review B\/} {\bf 107} 184439

\bibitem{xue2020control}
Xue F, Wang Z, Hou Y, Gu L and Wu R 2020 {\em Physical Review B\/} {\bf 101} 184426

\bibitem{kim2020observation}
Kim J, Son S, Coak M~J, Hwang I, Lee Y, Zhang K and Park J~G 2020 {\em Journal of Applied Physics\/} {\bf 128}

\bibitem{behura2021moire}
Behura S~K, Miranda A, Nayak S, Johnson K, Das P and Pradhan N~R 2021 {\em Emergent Materials\/} {\bf 4} 813--826

\bibitem{tong2019magnetic}
Tong Q, Chen M and Yao W 2019 {\em Physical Review Applied\/} {\bf 12} 024031

\bibitem{fazekas1999lecture}
Fazekas P 1999 {\em Lecture notes on electron correlation and magnetism\/} vol~5 (World scientific)

\bibitem{li2022free}
Li X, Jiang S, Meng Q, Zuo H, Zhu Z, Balents L and Behnia K 2022 {\em Physical Review B\/} {\bf 106} L020402

\bibitem{huang2017layer}
Huang B, Clark G, Navarro-Moratalla E, Klein D~R, Cheng R, Seyler K~L, Zhong D, Schmidgall E, McGuire M~A, Cobden D~H {\em et~al.\/} 2017 {\em Nature\/} {\bf 546} 270--273

\bibitem{chen2019direct}
Chen W, Sun Z, Wang Z, Gu L, Xu X, Wu S and Gao C 2019 {\em Science\/} {\bf 366} 983--987

\bibitem{liu2016exfoliating}
Liu J, Sun Q, Kawazoe Y and Jena P 2016 {\em Physical Chemistry Chemical Physics\/} {\bf 18} 8777--8784

\bibitem{wu2019strain}
Wu Z, Yu J and Yuan S 2019 {\em Physical Chemistry Chemical Physics\/} {\bf 21} 7750--7755

\bibitem{webster2018strain}
Webster L and Yan J~A 2018 {\em Physical Review B\/} {\bf 98} 144411

\bibitem{xie2023evidence}
Xie H, Luo X, Ye Z, Sun Z, Ye G, Sung S~H, Ge H, Yan S, Fu Y, Tian S {\em et~al.\/} 2023 {\em Nature Physics\/} {\bf 19} 1150--1155

\bibitem{li2019pressure}
Li T, Jiang S, Sivadas N, Wang Z, Xu Y, Weber D, Goldberger J~E, Watanabe K, Taniguchi T, Fennie C~J {\em et~al.\/} 2019 {\em Nature materials\/} {\bf 18} 1303--1308

\bibitem{soriano2019interplay}
Soriano D, Cardoso C and Fern{\'a}ndez-Rossier J 2019 {\em Solid State Communications\/} {\bf 299} 113662

\bibitem{song2021direct}
Song T, Sun Q~C, Anderson E, Wang C, Qian J, Taniguchi T, Watanabe K, McGuire M~A, St{\"o}hr R, Xiao D {\em et~al.\/} 2021 {\em Science\/} {\bf 374} 1140--1144

\bibitem{sivadas2018stacking}
Sivadas N, Okamoto S, Xu X, Fennie C~J and Xiao D 2018 {\em Nano letters\/} {\bf 18} 7658--7664

\bibitem{jiang2019stacking}
Jiang P, Wang C, Chen D, Zhong Z, Yuan Z, Lu Z~Y and Ji W 2019 {\em Physical Review B\/} {\bf 99} 144401

\bibitem{song2019switching}
Song T, Fei Z, Yankowitz M, Lin Z, Jiang Q, Hwangbo K, Zhang Q, Sun B, Taniguchi T, Watanabe K {\em et~al.\/} 2019 {\em Nature materials\/} {\bf 18} 1298--1302

\bibitem{xu2022coexisting}
Xu Y, Ray A, Shao Y~T, Jiang S, Lee K, Weber D, Goldberger J~E, Watanabe K, Taniguchi T, Muller D~A {\em et~al.\/} 2022 {\em Nature nanotechnology\/} {\bf 17} 143--147

\bibitem{yang2024macroscopic}
Yang B, Patel T, Cheng M, Pichugin K, Tian L, Sherlekar N, Yan S, Fu Y, Tian S, Lei H {\em et~al.\/} 2024 {\em Nature Communications\/} {\bf 15} 4982

\bibitem{huang2018electrical}
Huang B, Clark G, Klein D~R, MacNeill D, Navarro-Moratalla E, Seyler K~L, Wilson N, McGuire M~A, Cobden D~H, Xiao D {\em et~al.\/} 2018 {\em Nature nanotechnology\/} {\bf 13} 544--548

\bibitem{xie2022twist}
Xie H, Luo X, Ye G, Ye Z, Ge H, Sung S~H, Rennich E, Yan S, Fu Y, Tian S {\em et~al.\/} 2022 {\em Nature Physics\/} {\bf 18} 30--36

\bibitem{cheng2023electrically}
Cheng G, Rahman M~M, Allcca A~L, Rustagi A, Liu X, Liu L, Fu L, Zhu Y, Mao Z, Watanabe K {\em et~al.\/} 2023 {\em Nature Electronics\/} {\bf 6} 434--442

\bibitem{liu2018three}
Liu Y and Petrovic C 2018 {\em Physical Review B\/} {\bf 97} 014420

\bibitem{hauser1969magnetic}
Hauser J 1969 {\em Physical Review\/} {\bf 187} 580

\bibitem{bora2021magnetic}
Bora M and Deb P 2021 {\em Journal of Physics: Materials\/} {\bf 4} 034014

\bibitem{cardoso2023strong}
Cardoso C, Costa A, MacDonald A and Fern{\'a}ndez-Rossier J 2023 {\em Physical Review B\/} {\bf 108} 184423

\bibitem{zhong2017van}
Zhong D, Seyler K~L, Linpeng X, Cheng R, Sivadas N, Huang B, Schmidgall E, Taniguchi T, Watanabe K, McGuire M~A {\em et~al.\/} 2017 {\em Science advances\/} {\bf 3} e1603113

\bibitem{seyler2018valley}
Seyler K~L, Zhong D, Huang B, Linpeng X, Wilson N~P, Taniguchi T, Watanabe K, Yao W, Xiao D, McGuire M~A {\em et~al.\/} 2018 {\em Nano letters\/} {\bf 18} 3823--3828

\bibitem{ren2024spin}
Ren M~X, Zhang Y~J, Gao Y~M, Tian M~Y, Jin C~D, Zhang H, Lian R~Q, Gong P~L, Wang R~N, Wang J~L {\em et~al.\/} 2024 {\em Physical Review B\/} {\bf 109} 045420

\bibitem{parkin1993origin}
Parkin S 1993 {\em Physical review letters\/} {\bf 71} 1641

\bibitem{schneeloch2024role}
Schneeloch J~A, Aczel A~A, Ye F and Louca D 2024 {\em Physical Review B\/} {\bf 110} 144439

\bibitem{anderson2023programming}
Anderson E, Fan F~R, Cai J, Holtzmann W, Taniguchi T, Watanabe K, Xiao D, Yao W and Xu X 2023 {\em Science\/} {\bf 381} 325--330

\bibitem{shindou2013chiral}
Shindou R, Ohe J~i, Matsumoto R, Murakami S and Saitoh E 2013 {\em Physical Review B—Condensed Matter and Materials Physics\/} {\bf 87} 174402

\bibitem{tapia2024chiral}
Tapia I, Cazor X and Mellado P 2024 {\em Advanced Physics Research\/} {\bf 3} 2300135

\bibitem{ding2021field}
Ding L, Xu X, Jeschke H~O, Bai X, Feng E, Alemayehu A~S, Kim J, Huang F~T, Zhang Q, Ding X {\em et~al.\/} 2021 {\em Nature communications\/} {\bf 12} 5339

\bibitem{zimmermann2014ferroic}
Zimmermann A~S, Meier D and Fiebig M 2014 {\em Nature communications\/} {\bf 5} 4796

\bibitem{begum2022magnetic}
Begum~Popy R, Frank J and Stamps R~L 2022 {\em Journal of Applied Physics\/} {\bf 132}

\bibitem{kennes2021moire}
Kennes D~M, Claassen M, Xian L, Georges A, Millis A~J, Hone J, Dean C~R, Basov D, Pasupathy A~N and Rubio A 2021 {\em Nature Physics\/} {\bf 17} 155--163

\bibitem{gonzalez2019cold}
Gonz{\'a}lez-Tudela A and Cirac J~I 2019 {\em Physical Review A\/} {\bf 100} 053604

\bibitem{ganguli2023visualization}
Ganguli S~C, Aapro M, Kezilebieke S, Amini M, Lado J~L and Liljeroth P 2023 {\em Nano Letters\/} {\bf 23} 3412--3417

\bibitem{edelberg2020tunable}
Edelberg D, Kumar H, Shenoy V, Ochoa H and Pasupathy A~N 2020 {\em Nature Physics\/} {\bf 16} 1097--1102

\bibitem{abdul2021domain}
Abdul-Wahab D, Iacocca E, Evans R~F, Bedoya-Pinto A, Parkin S, Novoselov K~S and Santos E~J 2021 {\em Applied Physics Reviews\/} {\bf 8}

\bibitem{bhowmik2024higher}
Bhowmik S, Banerjee S and Saha A 2024 {\em Physical Review B\/} {\bf 109} 104417

\bibitem{li2022higher}
Li Y~M, Wu Y~J, Luo X~W, Huang Y and Chang K 2022 {\em Physical Review B\/} {\bf 106} 054403

\bibitem{mook2021chiral}
Mook A, D{\'\i}az S~A, Klinovaja J and Loss D 2021 {\em Physical Review B\/} {\bf 104} 024406

\bibitem{mcclarty2022topological}
McClarty P~A 2022 {\em Annual Review of Condensed Matter Physics\/} {\bf 13} 171--190

\bibitem{zhuo2025topological}
Zhuo F, Kang J, Manchon A and Cheng Z 2025 {\em Advanced Physics Research\/} {\bf 4} 2300054

\bibitem{kuepferling2023measuring}
Kuepferling M, Casiraghi A, Soares G, Durin G, Garcia-Sanchez F, Chen L, Back C~H, Marrows C~H, Tacchi S and Carlotti G 2023 {\em Reviews of Modern Physics\/} {\bf 95} 015003

\bibitem{hua2023magnon}
Hua C~B, Xiao F, Liu Z~R, Sun J~H, Gao J~H, Chen C~Z, Tong Q, Zhou B and Xu D~H 2023 {\em Physical Review B\/} {\bf 107} L020404

\bibitem{sil2020first}
Sil A and Ghosh A~K 2020 {\em Journal of Physics: Condensed Matter\/} {\bf 32} 205601

\bibitem{hirosawa2020magnonic}
Hirosawa T, Diaz S~A, Klinovaja J and Loss D 2020 {\em Physical review letters\/} {\bf 125} 207204

\bibitem{wang2023observation}
Wang H, Madami M, Chen J, Jia H, Zhang Y, Yuan R, Wang Y, He W, Sheng L, Zhang Y {\em et~al.\/} 2023 {\em Physical Review X\/} {\bf 13} 021016

\bibitem{lucassen2019tuning}
Lucassen J, Meijer M~J, Kurnosikov O, Swagten H~J, Koopmans B, Lavrijsen R, Kloodt-Twesten F, Fr{\"o}mter R and Duine R~A 2019 {\em Physical review letters\/} {\bf 123} 157201

\bibitem{finocchio2016magnetic}
Finocchio G, B{\"u}ttner F, Tomasello R, Carpentieri M and Kl{\"a}ui M 2016 {\em Journal of Physics D: Applied Physics\/} {\bf 49} 423001

\bibitem{kim2024emergence}
Kim K~M and Kim S~K 2024 {\em arXiv preprint arXiv:2408.05616\/}

\bibitem{akram2021moire}
Akram M, LaBollita H, Dey D, Kapeghian J, Erten O and Botana A~S 2021 {\em Nano Letters\/} {\bf 21} 6633--6639

\bibitem{fumega2023moire}
Fumega A~O and Lado J~L 2023 {\em 2D Materials\/} {\bf 10} 025026

\bibitem{huang2022ferroelectric}
Huang K, Shao D~F and Tsymbal E~Y 2022 {\em Nano Letters\/} {\bf 22} 3349--3355

\bibitem{sun2023theoretical}
Sun W, Wang W, Hu R, Yang C, Li L, Huang S, Li X and Cheng Z 2023 {\em ACS Applied Nano Materials\/} {\bf 6} 17021--17030

\bibitem{song2022evidence}
Song Q, Occhialini C~A, Erge{\c{c}}en E, Ilyas B, Amoroso D, Barone P, Kapeghian J, Watanabe K, Taniguchi T, Botana A~S {\em et~al.\/} 2022 {\em Nature\/} {\bf 602} 601--605

\bibitem{antao2024electric}
Ant{\~a}o T~V, Lado J~L and Fumega A~O 2024 {\em Nano Letters\/} {\bf 24} 15767--15773

\bibitem{jin2024skyrmion}
Jin Z, Zeng Z, Cao Y and Yan P 2024 {\em Physical Review Letters\/} {\bf 133} 196701

\bibitem{zhang2016magnetic}
Zhang X, Zhou Y and Ezawa M 2016 {\em Nature communications\/} {\bf 7} 10293

\bibitem{jiang2017direct}
Jiang W, Zhang X, Yu G, Zhang W, Wang X, Benjamin~Jungfleisch M, Pearson J~E, Cheng X, Heinonen O, Wang K~L {\em et~al.\/} 2017 {\em Nature Physics\/} {\bf 13} 162--169

\bibitem{lee1955magnetostriction}
Lee E~W 1955 {\em Reports on progress in physics\/} {\bf 18} 184

\bibitem{scholes2006excitons}
Scholes G~D and Rumbles G 2006 {\em Nature materials\/} {\bf 5} 683--696

\bibitem{sell1967optical}
Sell D~D, Greene R and White R~M 1967 {\em Physical Review\/} {\bf 158} 489

\bibitem{klaproth2023origin}
Klaproth T, Aswartham S, Shemerliuk Y, Selter S, Janson O, van~den Brink J, B{\"u}chner B, Knupfer M, Pazek S, Mikhailova D {\em et~al.\/} 2023 {\em Physical Review Letters\/} {\bf 131} 256504

\bibitem{diederich2023tunable}
Diederich G~M, Cenker J, Ren Y, Fonseca J, Chica D~G, Bae Y~J, Zhu X, Roy X, Cao T, Xiao D {\em et~al.\/} 2023 {\em Nature Nanotechnology\/} {\bf 18} 23--28

\bibitem{freeman1968exciton}
Freeman S and Hopfield J 1968 {\em Physical Review Letters\/} {\bf 21} 910

\bibitem{wang2023exciton}
Wang Z, Zhang X~X, Shiomi Y, Arima T~h, Nagaosa N, Tokura Y and Ogawa N 2023 {\em Physical Review Research\/} {\bf 5} L042032

\bibitem{wang2022light}
Wang X, Xiao C, Park H, Zhu J, Wang C, Taniguchi T, Watanabe K, Yan J, Xiao D, Gamelin D~R {\em et~al.\/} 2022 {\em Nature\/} {\bf 604} 468--473

\bibitem{zhang2022magnon}
Zhang Y, Kim H, Zhang W, Watanabe K, Taniguchi T, Gao Y, Maruyama M, Okada S, Shinokita K and Matsuda K 2022 {\em Advanced Materials\/} {\bf 34} 2200301

\bibitem{ziebel2024crsbr}
Ziebel M~E, Feuer M~L, Cox J, Zhu X, Dean C~R and Roy X 2024 {\em Nano Letters\/} {\bf 24} 4319--4329

\bibitem{liu2024intralayer}
Liu N, Wang C, Zhang Y, Pang F, Cheng Z, Zhang Y and Ji W 2024 {\em Physical Review B\/} {\bf 109} 214422

\bibitem{shi2024giant}
Shi J, Wang D, Jiang N, Xin Z, Zheng H, Shen C, Zhang X and Liu X 2024 {\em arXiv preprint arXiv:2409.18437\/}

\bibitem{bae2022exciton}
Bae Y~J, Wang J, Scheie A, Xu J, Chica D~G, Diederich G~M, Cenker J, Ziebel M~E, Bai Y, Ren H {\em et~al.\/} 2022 {\em Nature\/} {\bf 609} 282--286

\bibitem{cenker2022reversible}
Cenker J, Sivakumar S, Xie K, Miller A, Thijssen P, Liu Z, Dismukes A, Fonseca J, Anderson E, Zhu X {\em et~al.\/} 2022 {\em Nature Nanotechnology\/} {\bf 17} 256--261

\bibitem{jin2019observation}
Jin C, Regan E~C, Yan A, Iqbal Bakti~Utama M, Wang D, Zhao S, Qin Y, Yang S, Zheng Z, Shi S {\em et~al.\/} 2019 {\em Nature\/} {\bf 567} 76--80

\bibitem{das2025moire}
Das J and Erten O 2025 {\em Physical Review B\/} {\bf 111} 184406

\bibitem{regan2024spin}
Regan E~C, Lu Z, Wang D, Zhang Y, Devakul T, Nie J~H, Zhang Z, Zhao W, Watanabe K, Taniguchi T {\em et~al.\/} 2024 {\em Nature communications\/} {\bf 15} 10252

\bibitem{akif2024theory}
Akif~Keskiner M, Ghaemi P, Oktel M~O and Erten O 2024 {\em Nano Letters\/} {\bf 24} 8575--8579

\bibitem{nagaosa2010anomalous}
Nagaosa N, Sinova J, Onoda S, MacDonald A~H and Ong N~P 2010 {\em Reviews of modern physics\/} {\bf 82} 1539--1592

\bibitem{tschirhart2023intrinsic}
Tschirhart C, Redekop E, Li L, Li T, Jiang S, Arp T, Sheekey O, Taniguchi T, Watanabe K, Huber M {\em et~al.\/} 2023 {\em Nature Physics\/} {\bf 19} 807--813

\bibitem{zhou2025itinerant}
Zhou W~T, Dong Z~Y, Gu Z~L and Li J~X 2025 {\em arXiv preprint arXiv:2502.10991\/}

\bibitem{zhang2014proximity}
Zhang J, Triola C and Rossi E 2014 {\em Physical review letters\/} {\bf 112} 096802

\bibitem{gong2024spin}
Gong Z, Zhang Q~Q, Mu H~Y, An X~T and Liu J~J 2024 {\em Physical Review B\/} {\bf 109} 045301

\bibitem{PhysRevX.12.031042}
\ifmmode~\check{S}\else \v{S}\fi{}mejkal L, Sinova J and Jungwirth T 2022 {\em Phys. Rev. X\/} {\bf 12}(3) 031042 \urlprefix\url{https://link.aps.org/doi/10.1103/PhysRevX.12.031042}

\bibitem{mazin2022altermagnetism}
Mazin I and editors P 2022 Altermagnetism—a new punch line of fundamental magnetism

\bibitem{sheoran2024nonrelativistic}
Sheoran S and Bhattacharya S 2024 {\em Physical Review Materials\/} {\bf 8} L051401

\bibitem{liu2024twisted}
Liu Y, Yu J and Liu C~C 2024 {\em Physical Review Letters\/} {\bf 133} 206702

\bibitem{coleman1995simple}
Coleman P, Ioffe L and Tsvelik A~M 1995 {\em Physical Review B\/} {\bf 52} 6611

\bibitem{guerci2024topological}
Guerci D, Lucht K~P, Cr{\'e}pel V, Cano J, Pixley J and Millis A 2024 {\em Physical Review B\/} {\bf 110} 165128

\bibitem{zhao2023gate}
Zhao W, Shen B, Tao Z, Han Z, Kang K, Watanabe K, Taniguchi T, Mak K~F and Shan J 2023 {\em Nature\/} {\bf 616} 61--65

\bibitem{zhao2024emergence}
Zhao W, Shen B, Tao Z, Kim S, Kn{\"u}ppel P, Han Z, Zhang Y, Watanabe K, Taniguchi T, Chowdhury D {\em et~al.\/} 2024 {\em Nature Physics\/} {\bf 20} 1772--1777

\bibitem{xie2024kondo}
Xie F, Chen L and Si Q 2024 {\em Physical Review Research\/} {\bf 6} 013219

\bibitem{leeb2021anomalous}
Leeb V, Polyudov K, Mashhadi S, Biswas S, Valent{\'\i} R, Burghard M and Knolle J 2021 {\em Physical Review Letters\/} {\bf 126} 097201

\bibitem{takagi2019concept}
Takagi H, Takayama T, Jackeli G, Khaliullin G and Nagler S~E 2019 {\em Nature Reviews Physics\/} {\bf 1} 264--280

\bibitem{liu2021towards}
Liu H 2021 {\em International Journal of Modern Physics B\/} {\bf 35} 2130006

\bibitem{ge2022enhanced}
Ge M, Wang H, Wu J, Si C, Zhang J and Zhang S 2022 {\em npj Computational Materials\/} {\bf 8} 32

\bibitem{he2020giant}
He W~Y, Goldhaber-Gordon D and Law K~T 2020 {\em Nature communications\/} {\bf 11} 1650

\bibitem{wang2024fractional}
Wang C, Zhang X~W, Liu X, He Y, Xu X, Ran Y, Cao T and Xiao D 2024 {\em Physical Review Letters\/} {\bf 132} 036501

\bibitem{farooq2019switchable}
Farooq M~U and Hong J 2019 {\em npj 2D Materials and Applications\/} {\bf 3} 3

\bibitem{liu2020bandgap}
Liu X, Song C, Wu Z, Wang J, Pan J and Li C 2020 {\em Journal of Physics D: Applied Physics\/} {\bf 53} 385002

\bibitem{pawbake2022high}
Pawbake A, Pelini T, Delhomme A, Romanin D, Vaclavkova D, Martinez G, Calandra M, Measson M~A, Veis M, Potemski M {\em et~al.\/} 2022 {\em ACS nano\/} {\bf 16} 12656--12665

\bibitem{maity2025electron}
Maity S, Das S, Palit M, Dey K, Das B, Kundu T, Paramanik R, De B~K, Kunwar H~S and Datta S 2025 {\em Physical Review B\/} {\bf 111} L140407

\bibitem{li2019intrinsic}
Li J, Li Y, Du S, Wang Z, Gu B~L, Zhang S~C, He K, Duan W and Xu Y 2019 {\em Science advances\/} {\bf 5} eaaw5685

\bibitem{xiao2020chiral}
Xiao C, Tang J, Zhao P, Tong Q and Yao W 2020 {\em Physical Review B\/} {\bf 102} 125409

\bibitem{shaban2023skyrmion}
Shaban P, Lobanov I, Uzdin V and Iorsh I 2023 {\em Physical Review B\/} {\bf 108} 174440

\bibitem{balents2020superconductivity}
Balents L, Dean C~R, Efetov D~K and Young A~F 2020 {\em Nature Physics\/} {\bf 16} 725--733

\bibitem{adak2024tunable}
Adak P~C, Sinha S, Agarwal A and Deshmukh M~M 2024 {\em Nature Reviews Materials\/} {\bf 9} 481--498

\bibitem{kezilebieke2022moire}
Kezilebieke S, Vano V, Huda M~N, Aapro M, Ganguli S~C, Liljeroth P and Lado J~L 2022 {\em Nano letters\/} {\bf 22} 328--333

\end{thebibliography}
\end{document}